\documentclass[10pt]{article}
\usepackage{txfonts}
\usepackage{fancyhdr}
\usepackage{titlesec}
\usepackage{leftidx}
\usepackage{cite}
\usepackage{ifthen}

\titleformat{\section}{\centering\large\bfseries}{\S\arabic{section}}{1em}{}
\newboolean{first}
\setboolean{first}{true}

\begin{document}

\setlength\abovedisplayskip{2pt}
\setlength\abovedisplayshortskip{0pt}
\setlength\belowdisplayskip{2pt}
\setlength\belowdisplayshortskip{0pt}

\title{\bf \Large  A New Construction of Multi-receiver Authentication Codes from
Pseudo-Symplectic Geometry over Finite Fields
\author{ Xiuli Wang \\ \small \it \rm \emph{(College of Science, Civil Aviation
University of China, Tianjin, 300300, P.R.China.)} }\date{}}
\maketitle

{\small {\bf Abstract:}\hskip 2mm {\small Multi-receiver
authentication codes allow one sender to construct an authenticated
message for a group of receivers such that each receiver can verify
authenticity of the received message. In this paper, we constructed
one multi-receiver authentication codes from pseudo-symplectic
geometry over finite fields. The parameters and the probabilities of
deceptions of this codes are also computed.}

 }

{\bf 2000 MR Subject Classification}:\ {\rm 15A03; 94A60; 94A62}

\footnote{Supported by the NSF of China(61179026)and Fundamental
Research of the Central Universities of China Civil Aviation
University of Science special (ZXH2012k003).} \footnote{Address:
College of Science, Civil Aviation University of China, Tianjin
300300, P.R.China.} \footnote{E-mail: xlwang@cauc.edu.cn, \
wangxiuli1999@tom.com}

\thispagestyle{fancyplain} \fancyhead{}
\fancyhead[L]{\textit{2013: My paper}\\
} \fancyfoot{} \vskip 10mm

\begin{center}
{\large\bf\S\bf1\hspace*{3mm}  Introduction}
\end{center}

Multi-receiver authentication codes (MRA-codes) are introduced by
Desmedt, Frankel, and Yung (DFY) $^{[1]}$ as an extension of
Simmons' model of unconditionally secure authentication. In an
MRA-codes, a sender wants to authenticate a message for a group of
receivers such that each receiver can verify authenticity of the
received message. There are three phases in an MRA-codes:

1. $Key\ distribution$. The $KDC$ (key distribution centre)
privately transmits the key information to the sender and each
receiver (the sender can also be the $KDC$).

2. $Broadcast$. For a source state, the sender generates the
authenticated message using his/her key and broadcasts the
authenticated message.

3. $Verification$. Each user can verify the authenticity of the
broadcast message.

Denote by $X_{1}\times\cdots\times X_{n}$ the direct product of sets
$X_{1},\cdots,X_{n}$, and by $p_i$ the projection mapping of
$X_{1}\times\cdots\times X_{n}$ on $X_i$. That is, $p_i :
X_{1}\times\cdots\times X_{n}\rightarrow X_i$ defined by $p_i(x_1,
x_2,\cdots,x_n)=x_i$. Let $g_1:X_1\rightarrow Y_1$ and
$g_2:X_2\rightarrow Y_2$ be two mappings, we denote the direct
product of $g_1$ and $g_2$ by $g_1\times g_2$, where $g_1\times g_2:
X_1\times X_2\rightarrow Y_1\times Y_2$ is defined by $(g_1\times
g_2)(x_1,x_2)=(g_1(x_1),g_2(x_2))$. The identity mapping on a set
$X$ is denoted by $1_X$.

Let $C=(S, M, E, f )$ and $C_i=(S, M_i , E_i , f_i), i=1, 2, ...,
n$, be authentication codes. We call $(C; C_1 , C_2 ,\cdots , C_n)$
a multi-receiver authentication code (MRA-code) if there exist two
mappings $\tau: E\rightarrow E_1\times\cdots\times E_n$ and $\pi:
M\rightarrow M_1\times\cdots\times Mn$ such that for any $(s, e)\in
S\times E$ and any $1\leq i\leq n$, the following identity holds
$$p_i(\pi f(s, e))=f_i ((1_S\times p_i\tau (s, e)).$$ Let $\tau_i=p_i\tau$ and
$\pi_i=p_i\pi$. Then we have for each $(s, e)\in S\times E$$$\pi_if
(s, e)= f_i(1_S\times \tau_i)(s, e).$$

We adopt Kerckhoff's principle that everything in the system except
the actual keys of the sender and receivers is public. This includes
the probability distribution of the source states and the sender's
keys.

$Attackers$ could be $outsiders$ who do not have access to any key
information, or $insiders$ who have some key information. We only
need to consider the latter group as it is at least as powerful as
the former. We consider the systems that protect against the
coalition of groups of up to a maximum size of receivers, and we
study impersonation and substitution attacks.

Assume there are $n$ receivers $R_{1},\cdots,R_{n}.$ Let
$L=\{i_{1},\cdots,i_{l}\}\subseteq\{1,\cdots,n\},
R_{L}=\{R_{i_{1}},\cdots,R_{i_{l}}\}$ and
$E_{L}=E_{R_{i_{1}}}\times\cdots\times E_{R_{i_{l}}}$. We consider
the attack from $R_{L}$ on a receiver $R_{i}$, where $i\notin L$.

$Impersonation\ attack$: $R_L$, after receiving their secret keys,
send a message $m$ to $R_i$. $R_L$ is successful if $m$ is accepted
by $R_i$ as authentic. We denote by $P_I[i, L]$ the success
probability of $R_L$ in performing an impersonation attack on $R_i$.
This can be expressed as $$P_I[i, L]=\max\limits_{e_L\in
E_L}\max\limits_{m\in M} P(m\ {\rm is\ accepted\ by}\ R_i | e_L)$$
where $i\notin L$.

$Substitution\ attack$: $R_L$, after observing a message $m$ that is
transmitted by the sender, replace $m$ with another message $m'$.
$R_L$ is successful if $m'$ is accepted by $R_i$ as authentic. We
denote by $P_S[i, L]$ the success probability of $R_L$ in performing
a substitution attack on $R_i$ . We have $$P_S[i,L]=\max\limits_{e_L
\in E_L}\max\limits_{m\in M}\max\limits_{m'\neq m\in M} P(R_i\ {\rm
accepts}\ m' | m, e_L)$$  where $i\notin L$.

\begin{center}
{\large\bf\S\bf2\hspace*{3mm} Pseudo-Symplectic Geometry}
\end{center}

Let $ F_q$ be the finite field with $q$ elements, where $q$ is a
power of $2$, $n=2\nu + \delta$ and $\delta$=1,2. Let
$$
  K = \left (
  \begin {array}{ccc}
  0 & I^{(\nu)}\\
  I^{(\nu)}& 0
  \end{array} \right ), \,\,\,\,\,  S_{1} = \left (
  \begin {array}{cc}
  K &\\
 & {1}
  \end{array} \right ),\,\,\,\,\,S_{2} = \left (
  \begin {array}{cccc}
   K &&\\
& 0&1\\
&1&1\end{array} \right )
  $$
and $S_\delta$ is an $(2\nu+\delta)\times(2\nu+\delta)$
non-alternate symmetric matrix.\

The pseudo-symplectic group of degree $(2\nu+\delta)$ over $ F_q$ is
defined to be the set of matrices $Ps_{2\nu + \delta}(F_q)=\{T|T
S_\delta\ ^{t}T=S_\delta\}$ denoted by $Ps_{2\nu + \delta}(F_q)$.\

Let $F_q^{(2\nu + \delta)}$ be the $(2\nu + \delta)$ -dimensional
row vector space over $F_q$. $Ps_{2\nu + \delta}(F_q)$ has an action
on $F_q^{(2\nu + \delta)}$ defined as follows
$$
F_q^{(2\nu + \delta)} \times Ps_{2\nu + \delta}(F_q) \rightarrow
F_q^{(2\nu + \delta)}\ $$ $$ ((x_1,x_2,\dots,x_{2\nu +
\delta}),T)\rightarrow(x_1,x_2,\dots,x_{2\nu + \delta})T.
$$
The vector space $F_q^{(2\nu + \delta)}$ together with this group
action is called the pseudo-symplectic space over the finite field $
F_q$ of characteristic 2.\

Let $P$ be an $m$-dimensional subspace of $F_q^{(2\nu + \delta)}$,
then $PS_\delta\ ^{t}P$ is cogredient to one of the following three
normal forms
$$ M(\,m,\,2s,\,s\,)=\left (
  \begin {array}{ccccc}
  0&I^{(s)}&\\
  I^{(s)}&0&\\
  &&0^{(m-2s)}
  \end{array} \right)$$\\
$$ M(\,m,\,2s+1,\,s\,)=\left (
  \begin {array}{ccccc}
  0&I^{(s)}&\\
  I^{(s)}&0&\\
   &&1\\
  &&&0^{(m-2s-1)}
  \end{array} \right)$$\\
$$ M(\,m,\,2s+2,\,s\,)=\left (
  \begin {array}{ccccc}
  0&I^{(s)}&\\
  I^{(s)}&0&\\
   &&0&1&\\
   &&1&1&\\
  &&&&0^{(m-2s-2)}
  \end{array} \right)$$\\
for some $s$ such that $0\leq s\leq [m/2]$. We say that $P$ is a
subspace of type $(m,2s+\tau,s,\epsilon)$, where $\tau$ =0,1 or 2
and $\epsilon$ =0 or 1, if \

(i) $P S_\delta\ ^{t}P$ is cogredient to $M(m,2s+\tau,s)$, and\

 (ii) $e_{2\nu +1}\notin P$ or $e_{2\nu +1}\in P$ according to $\epsilon=0$ or
 $\epsilon=1$, respectively.\

Let $P$ be an $m$-dimensional subspace of $F_q^{(2\nu + \delta)}$.
Denote by $P^\perp $ the set of vectors which are orthogonal to
every vector of $P$, i.e.,
$$ P^\perp=\{y\in F_q^{(2\nu +
\delta)}|yS_\delta\ ^{t}x=0\, for\,\, all\, x\in P\}.$$Obviously,
$P^\perp $ is a $(2\nu +\delta-m)$-dimensional subspace of
$F_q^{(2\nu + \delta)}$.

More properties of pseudo-symplectic geometry over finite fields
 can be found in [2].

In [3], Desmedt, Frankel and Yung gave two constructions for
MRA-codes based on polynomials and finite geometries, respectively.
There are other constructions of multi-receiver authentication codes
are given in $[4-7]$. The construction of authentication codes is
combinational design in its nature. We know that the geometry of
classical groups over finite fields, including symplectic geometry,
pseudo-symplectic geometry, unitary geometry and orthogonal geometry
can provide a better combination of structure and easy to count. In
this paper we constructed one multi-receiver authentication codes
from pseudo-symplectic geometry over finite fields. The parameters
and the probabilities of deceptions of this codes are also computed.
We realize the generalization of the results of the article [8] from
symplectic geometry to pseudo-symplectic geometry over Finite
Fields.

\begin{center}
{\large\bf\S\bf3\hspace*{3mm}  Construction}
\end{center}

Let $ \mathbb{F}_{q}$ be a finite field with $q$ elements and
$e_i(1\leq i \leq 2\nu+2)$ be the row vector in
$\mathbb{F}_{q}^{(2\nu+2)}$ whose $i-$th coordinate is 1 and all
other coordinates are 0. Assume that $2<n+1<r<\nu$. $U=\langle
e_{1},e_{2},\cdots, e_{n}\rangle $, i.e., $U$ is an $n-$dimensional
subspace of $\mathbb{F}_{q}^{(2\nu+2)}$ generated by
$e_{1},e_{2},\cdots, e_{n}$, then $U^\perp=\langle e_{1},\cdots,
e_{\nu},e_{\nu +n+1},\cdots, e_{2\nu+2}\rangle$. The set of source
states $S$=$\{s|s\ {\rm is\ a\ subspace\ of\ type}\
(2r-n+1,2(r-n),r-n,1)\ {\rm and}\ U\subset s\subset U^\perp\}$; the
set of transmitter's encoding rules $E_T$=$\{e_T |e_T$ is a subspace
of type $(2n,2n,n,0)$\ and\ $U\subset e_T\}$; the set of $i-th$
receiver's decoding rules $E_{R_i}$=$\{e_{R_i} |e_{R_i}$ is a
subspace of type $(n+1,0,0,0)$ which is orthogonal to $\langle
e_{1},\cdots,e_{i-1},e_{i+1},\cdots, e_{n}\rangle$\}, $1\leq i \leq
n$; the set of messages $M=\{m|m\ {\rm is\ a\ subspace\ of\ type}\
(2r+1,2r,r,1)\ {\rm and}\ U\subset m\}$.

1. $Key\ Distribution.$ The $KDC$ randomly chooses a subspace
$e_T\in E_T$, then privately sends $e_T$ to the sender $T$. Then
$KDC$ randomly chooses a subspace $e_{R_i}\in E_{R_i}$ and
$e_{R_i}\subset e_T$, then privately sends $e_{R_i}$ to the $i-th$
receiver, where $1\leq i\leq n$.

2. $Broadcast.$ For a source state $s\in S$, the sender calculates
$m=s+e_T$ and broadcast $m$.

3. $Verification.$ Since the receiver $R_i$ holds the decoding rule
$e_{R_i}$, $R_i$ accepts $m$ as authentic if $e_{R_i}\subset m$.
$R_i$ can get $s$ from $s=m\cap U^{\perp}$.

{\bf Lemma 3.1} \ The above construction of multi-receiver
authentication codes is reasonable, that is

(1) \  $s+e_{T}=m\in M$, for all $ s\in S$ and $e_{T}\in E_{T}$;

(2) \ for any $ m\in M$,  $s=m\cap U^{\perp}$ is the uniquely source
state contained in $m$ and there is $e_{T}\in E_{T}$, such that
$m=s+e_{T}$.

{\bf \emph{Proof.}} (1)  \ \ For any $ s\in S,$ $ e_{T}\in E_{T}$,
Because $s$ is a subspace of  type $(2r-n,2(r-n),r-n,1) \ {\rm and}\
U\subset s\subset U^\perp\}$, we can assume that
$$\begin {array}[t]
{rr@{\extracolsep{0.2ex}}r}s=\left (
\begin {array}{ccccc}
U\\
Q\\
e_{2\nu+1}
\end{array} \right)\begin {array}{l}
\scriptstyle n\\ \scriptstyle 2(r-n)\\\ \ \scriptstyle 1
\end{array}\\\begin {array}{ccc} \scriptstyle&\,\,\,\,\,\,\scriptstyle \end{array}&
\end{array}$$ and $$\begin {array}[t]
{cc@{\extracolsep{0.2ex}}c}{\left (\begin {array}{cccccc}
U\\
Q\\
e_{2\nu+1}
\end{array}\right)}S_2\leftidx{^t}{\left (\begin
{array}{cccccc}
U\\
Q\\
e_{2\nu+1}
\end{array}\right)}=\left (
\begin {array}{cccc}
0^{(n)}&0&0&0\\
0&0&I^{(r-n)}&0\\
0&I^{(r-n)}&0&0\\
0&0&\ 0&0
\end{array} \right)
\end{array},$$
$$\begin {array}[t] {cc@{\extracolsep{0.2ex}}c}e_T=\left(
\begin {array}{ccccc}
U\\
V\\
\end{array} \right)\begin {array}{l}
\scriptstyle n\\ \scriptstyle n
\end{array}\\&\begin {array}{cc} \scriptstyle&\,\,\,\,\,\,\scriptstyle \end{array}&
\end{array}$$ and
$$\begin {array}[t]
{cc@{\extracolsep{0.2ex}}c}{\left (\begin {array}{cccccc}
U\\
V
\end{array}\right)}S_2\leftidx{^t}{\left (\begin {array}{cccccc}
U\\
V\end{array}\right)}S_2\leftidx{^t}{\left (\begin {array}{cccccc}
U\\
V
\end{array}\right)}=\left (
\begin {array}{ccc}
0&I^{(n)}\\
I^{(n)}&0
\end{array} \right)
\end{array}.$$\\
Obviously, for any $v \in V$ and $v\not=0$,$v \notin s$, therefore,
$$\begin {array}[t] {rr@{\extracolsep{0.2ex}}r}m=s+e_T=\left (
\begin {array}{c}
U\\
V\\
Q\\
e_{2\nu+1}
\end{array} \right)
\\\begin {array}{rr}\ \ \ \ \ \ \ \scriptstyle &\scriptstyle \end{array}
\end{array},$$ and
$$\begin {array}[t]
{cc@{\extracolsep{0.2ex}}c}{\left (\begin {array}{cccccc}
U\\
V\\
Q\\
e_{2\nu+1}
\end{array}\right)}S_2\leftidx{^t}{\left (\begin {array}{cccccc}
U\\
V\\
Q\\
e_{2\nu+1}
\end{array}\right)}=\left (
\begin {array}{ccccc}
0&I^{(n)}&0\ \ &0\ \ &0\\
I^{(n)}&0&0\ \ &0\ \ &0\\
0&0&0&I^{(r-n)}\ &0\\
0&0&I^{(r-n)}&\ 0\ \ &0\\
0&0&0&\ \ 0\ \ &0\\
\end{array} \right)
\end{array}.$$\\
From above, $m$ is a subspace of type $(2r+1,2r,r,1)$ and $U\subset
m$, i.e., $m\in M$.

(2) \ \ For $m\in M$, $m$ is a subspace of type $(2r+1,2r,r,1)$ and
$U \subset m$, so there is a subspace $V \subset m$, satisfying
$$\begin {array}[t]
{cc@{\extracolsep{0.2ex}}c}{\left (\begin {array}{cccccc}
U\\
V\end{array}\right)}S_2\leftidx{^t}{\left (\begin {array}{cccccc}
U\\
V
\end{array}\right)}=\left (
  \begin {array}{ccc}
0&I^{(n)}\\
I^{(n)}&0
\end{array} \right)
\end{array}.$$
Then we can assume that $$\begin {array}[t]
{rr@{\extracolsep{0.2ex}}r}m=\left (
\begin {array}{c}
U\\
V\\
Q\\
e_{2\nu+1}
\end{array} \right)
\\\begin {array}{rr}\ \ \ \ \ \ \ \scriptstyle &\scriptstyle \end{array}
\end{array}.$$
and satisfying
$$\begin {array}[t]
{cc@{\extracolsep{0.2ex}}c}{\left (\begin {array}{cccccc}
U\\
V\\
Q\\
e_{2\nu+1}
\end{array}\right)}S_2\leftidx{^t}{\left (\begin {array}{cccccc}
U\\
V\\
Q\\
e_{2\nu+1}
\end{array}\right)}=\left (
\begin {array}{ccccc}
0&I^{(n)}&0\ \ &0\ \ &0\\
I^{(n)}&0&0\ \ &0\ \ &0\\
0&0&0&I^{(r-n)}\ \ &0\\
0&0&I^{(r-n)}\ \ 0\ \ &0\\
0&0&0&0\ \ &0
\end{array} \right)
\end{array}.$$\\
Let $$s=\left(
\begin {array}{c}
U\\
Q\\
e_{2\nu+1}
\end{array} \right).$$\\
For $s\ {\rm is\ a\ subspace\ of\ type}\ (2r-n+1,2(r-n),r-n,1)\ {\rm
and}\ U\subset s\subset U^{\perp}$, i.e., $s\in S$ is a source
state. For any $v \in V$ and $v\not=0$,$v \notin s$ is obvious,
i.e., $V\cap U^\perp=\{0\}$. Therefore, $m\cap U^\perp=\left (
\begin {array}{c}
U\\
Q\\
e_{2\nu+1}
\end{array} \right)=s.$ Let $e_T=\left (
\begin {array}{c}
U\\
V
\end{array} \right)$, then $e_T$ is a transmitter's encoding rule and
satisfying $m=s+e_T$.\\
If $s'$ is another source state contained in $m$, then $U \subset
s'\subset U^{\perp}$. Therefore, $s'\subset m \cap U^{\perp}=s$,
while dim$s'$=dim$s$, so $s'$=$s$, i.e., $s$ is the uniquely source
state contained in $m$.\

From Lemma 3.1, we know that such construction of multi-receiver
authentication codes is reasonable and there are $n$ receivers in
this system. Next we compute the parameters of this codes.

{\bf Lemma 3.2} \ The parameters of this construction are
$$|S|=N(2(r-n),2(r-n),r-n,0;2\nu+2);\ \
|E_T|=q^{n(\nu -n+1)};\ \  |E_{R_i}|=q^{\nu-n+1}.$$

{\bf \emph{Proof.}} Since $U\subset s\subset U^{\perp}$, $s$ has the
form as follows
$$s=\left (
\begin {array}{cccccc}
I^{(n)}&0&0&0&0&0\\
0&B_2&0&B_4&0&0\\
0&0&0&0&1&0
\end{array} \right),$$
where $B_2,B_4$ is a subspace of type $(2(r-n),2(r-n),r-n,0)$ in the
pseudo-symplectic space ${F_q}^{(2\nu+2)}$. So
$|S|=N(2(r-n),2(r-n),r-n,0;2\nu+2)$.

Since $e_T$ is a subspace of type $(2n,2n,n,0)$, $e_T$ has the form
as follows
$$\begin {array}[t] {cc@{\extracolsep{0.2ex}}c}e_T=\left (
\begin {array}{cccccc}
I^{(n)}&0&0&0&0&0\\
0&R_2&I^{(n)}&R_4&R_5&R_6
\end{array} \right)
\\\begin {array}{cccccc}\ \ \ \ \ \ \ \ \scriptstyle n\ \ &\scriptstyle \nu-n\ &\scriptstyle n
\ &\scriptstyle \nu-n\ \ &\scriptstyle 1\ \ &\scriptstyle
1\end{array}
\end{array}.$$ For $e_T$ is a subspace of type $(2n,2n,n,0)$,
so $R_4=0$ and $R_6=0$, $R_2,R_5$ arbitrarily. Therefore
$|E_T|=q^{n(\nu -n+1)}$.\\

For any $e_{R_i}\in E_{R_i}$, $e_{R_i}$ is a subspace of type
$(n+1,0,0,0)$ which is orthogonal to $\langle
e_{1},\cdots,e_{i-1},e_{i+1},\cdots, e_{n}\rangle$, $1\leq i \leq
n$. So we can assume that
$$\begin {array}[t] {cc@{\extracolsep{0.2ex}}c}e_{R_{i}}=\left(
\begin {array}{cccccccccc}
I^{(l)}&0&0\ &0&0&\ \ \ \ 0&\ 0&\ 0&0&0\\
0&I^{(n-l)}&0\ &0&0&\ \ \ \ 0&\ 0&\ 0&0&0\\
0&0&H'_3\ &0&0&\ \ \ \ 1&\ 0&\ H'_8&H'_9&H'_{10}
\end{array} \right)\begin {array}{c}\end{array}
\\\begin {array}{cccccccccc}\ \ \ \ \ \ \ \ \scriptstyle l\ \ \ &\scriptstyle n-l\ \ \ &\scriptstyle
\nu-n\ &\scriptstyle l&\scriptstyle i-l-1\ &\scriptstyle 1\
&\scriptstyle n-i&\scriptstyle \nu-n\ &\scriptstyle 1\ \ \
&\scriptstyle 1\end{array}
\end{array}.$$ Since $e_{R_i}$ is a subspace of type
$(n+1,0,0,0),$ so $H'_8=0$ and $H'_{10}=0$, $H'_3, H'_9$
arbitrarily. Therefore, $|E_{R_i}|=q^{\nu-n+1}$.

{\bf Lemma 3.3} (1) The number of $e_T$ contained in $m$ is
$q^{n(r-n+1)}$;

(2) The number of the messages is
$|M|=q^{2n(\nu-r+1)}N(2(r-n),2(r-n),r-n,1;2\nu+2)$.

{\bf \emph{Proof.}} Let $m$ be a message, from the definition of
$m$, we may take $m$ as follows
$$\begin {array}[t]
{cc@{\extracolsep{0.2ex}}c}m=\left(
  \begin {array}{cccccccc}
I^{(n)}&0&0&0&0&0&0&0\\
0&I^{(r-n)}&0&0&0&0&0&0\\
0&0&0&I^{(n)}&0&0&0&0\\
0&0&0&0&I^{(r-n)}&0&\ 0&0\\
0&0&0&0&0&\ 0&\ 1&0
\end{array} \right)\begin {array}{c}\end{array}
\\\begin {array}{cccccccc}\ \ \ \ \ \ \ \ \ \scriptstyle n\ \ \ &\scriptstyle r-n\ &\scriptstyle
\nu-r\ &\scriptstyle n\ \ \ &\scriptstyle r-n&\ \ \ \scriptstyle
\nu-r&\scriptstyle 1\ &\scriptstyle 1\end{array}
\end{array}.$$\\
if $e_T \subset m$, then we can assume that
$$\begin {array}[t]
{cc@{\extracolsep{0.2ex}}c}e_T=\left(
\begin {array}{cccccccc}
I^{(n)}&0\ &0&0\ &0\ &0&0&0\\
0&R_2\ &0&I^{(n)}\ &0\ &0&R_7&0
\end{array} \right)\begin {array}{c}\end{array}
\\\begin {array}{cccccccc}\ \ \ \ \ \ \ \ \ \scriptstyle n\ &\scriptstyle r-n\ &\scriptstyle
\nu-r\ &\scriptstyle n\ &\scriptstyle r-n&\scriptstyle
\nu-r&\scriptstyle 1&\scriptstyle 1\end{array}
\end{array},$$\\
where $R_2$ and $R_7$ is arbitrarily. Therefore the number of $e_T$
which contained $m$ is $q^{n(r-n+1)}$;\\

(2) We know that a message contains only one source state and the
number of the transmitter's encoding rules contained in a message is
$q^{n(r-n+1)}$. Therefore we have
$|M|=|S||E_T|/q^{n(r-n+1)}=q^{n(\nu-r)}N(2(r-n),2(r-n),r-n,0;2\nu+2)$

Assume there are $n$ receivers $R_{1},\cdots,R_{n}.$ Let
$L=\{i_{1},\cdots,i_{l}\}\subseteq\{1,\cdots,n\},
R_{L}=\{R_{i_{1}},\cdots,R_{i_{l}}\}$ and
$E_{L}=E_{R_{i_{1}}}\times\cdots\times E_{R_{i_{l}}}$. We consider
the $impersonation\ attack$ and $substitution\ attack$ from $R_{L}$
on a receiver $R_{i}$, where $i\notin L$.

Without loss of generality, we can assume that
$R_{L}=\{R_{1},\cdots, R_{l}\},\ E_{L}=E_{R_{1}}\times\cdots\times
E_{R_{l}}$, where $1\leq l\leq n-1$. First, we will proof the
following results:

{\bf Lemma 3.4}\ For any $e_{L}=(e_{R_{1}},\cdots,e_{R_{l}})\in
E_{L}$, the number of $e_{T}$ containing $e_{L}$ is
$q^{(\nu-n+1)(n-l)}.$

\textbf{\emph{Proof.}}\ \ For any $e_{L}=(e_{R_{1}},\cdots,e_{R_{l}})\in E_{L}$, we
can assume that
$$\begin {array}[t] {cc@{\extracolsep{0.2ex}}c}e_L=\left(
  \begin {array}{cccccccccccc}
I^{(l)}\ \ &0&0&0&0\ &0&0&0\\
0&I^{(n-l)}\ \ &0&0&0\ &0&0&0\\
0&0&R_3&I^{(l)}\ \ &0\ &0&R_7&0
\end{array} \right)\begin {array}{c}\end{array}
\\\begin {array}{cccccccccc}\ \ \ \ \ \ \ \ \ \ \scriptstyle l\ \ \ \ \ &\scriptstyle n-l\ \ &\scriptstyle
\nu-n\ \ &\scriptstyle l\ \ \ &\scriptstyle n-l&\scriptstyle
\nu-n&\scriptstyle 1 &\scriptstyle 1\end{array}
\end{array}.$$\\
Therefore, $e_T$ containing $e_{L}$ has the form as follows
$$\begin {array}[t] {cc@{\extracolsep{0.2ex}}c}e_T=\left(
  \begin {array}{cccccccccccc}
I^{(l)}\ &0&0\ &0&0 &0&0&0\\
0\ &I^{(n-l)}&0\ &0&0&0&0&0\\
0\ &0&R_3\ &I^{(l)}&0&0&R_7&0\\
0\ &0&H_3\ &0&I^{(n-l)}&0&H_7&0
\end{array} \right)\begin {array}{c}\end{array}
\\\begin {array}{cccccccccc}\ \ \ \ \ \ \ \ \ \scriptstyle l\ \ \ \ \ &\scriptstyle n-l\ &\scriptstyle
\nu-n\ \ \ &\scriptstyle l\ \ \ \ &\scriptstyle n-l&\scriptstyle
\nu-n\ &\scriptstyle 1 \ &\scriptstyle 1\end{array}
\end{array},$$
where $H_3, H_7$ arbitrarily. Therefore, the number of $e_T$
containing $e_L$ is $q^{(\nu-n+1)(n-l)}.$

{\bf Lemma 3.5}\ For any $m\in M$ and $e_{L},e_{R_{i}}\subset m$,

(1) the number of $e_T$ contained in $m$ and containing $e_L$ is
$q^{(r-n+1)(n-l)}.$;

(2) the number of $e_T$ contained in $m$ and containing
$e_L,e_{R_{i}}$ is $q^{(n-l-1)(r-n+1)}$.

\textbf{\emph{Proof.}}\ \ (1) From the definition of $m$, we may
take $m$ as follows
$$\begin {array}[t]
{cc@{\extracolsep{0.2ex}}c}m=\left(
\begin {array}{cccccccccc}
I^{(l)}&0&0&0&0&0&0&0&0&0\\
0&I^{(n-l)}&0&0&0&0&0&0&0&0\\
0&0&I^{(r-n)}&0&0&0&0&0&0&0\\
0&0&0&0&I^{(l)}&0&0&0&0&0\\
0&0&0&0&0&I^{(n-l)}&0&0&0&0\\
0&0&0&0&0&0&I^{(r-n)}&0&0&0\\
0&0&0&0&0&0&0&0&1&0
\end{array} \right)\begin {array}{c}\end{array}
\\\begin {array}{cccccccccc}\ \ \ \ \ \ \ \ \ \ \scriptstyle l\ \ \ &\scriptstyle n-l\ \ \ &\scriptstyle r-n\ &\scriptstyle
\nu-r&\scriptstyle l\ \ \ &\scriptstyle n-l\ \ \ \ &\scriptstyle r-n
\ &\scriptstyle \nu-r&\scriptstyle 1&\scriptstyle 1\end{array}
\end{array}.$$

If $e_L \subset m$, then $e_L$ has the form as follows:

$$\begin {array}[t] {cc@{\extracolsep{0.2ex}}c}e_L=\left(
\begin {array}{cccccccccc}
I^{(l)}&0&0\ &0&0&0\ \ &0\ \ &0\ &0&0\\
0&I^{(n-l)}&0\ &0&0&0\ \ &0\ \ &0\ &0&0\\
0&0&R_3\ &0&I^{(l)}&0\ \ &0\ \ &0\ &R_9&0
\end{array} \right)\begin {array}{c}\end{array}
\\\begin {array}{cccccccccc}\ \ \ \ \ \ \ \ \ \scriptstyle l\ \ \ &\scriptstyle n-l\ &\scriptstyle r-n\ &\scriptstyle
\nu-r&\scriptstyle l\ &\scriptstyle n-l&\scriptstyle r-n &\
\scriptstyle \nu-r&\scriptstyle 1\ &\scriptstyle 1\end{array}
\end{array}.$$\\
If\ $e_{T}\subset m$ and $e_{T}\supset e_{L}$, then
$$\begin {array}[t] {cc@{\extracolsep{0.2ex}}c}e_T=\left(
\begin {array}{cccccccccc}
I^{(l)}&0&0\ &0&0&0&0\ &0&0&0\\
0&I^{(n-l)}&0\ &0&0&0&0\ &0&0&0\\
0&0&0\ &0&0&0&0\ &0&0&0\\
0&0&R_3\ &0&I^{(l)}&0&0\ &0&R_9&0\\
0&0&H_3\ &0&0&I^{(n-l)}&0\ &0&H_9&0
\end{array} \right)\begin {array}{c}\end{array}
\\\begin {array}{cccccccccc}\ \ \ \ \ \ \ \ \ \ \scriptstyle l\ \ \ &\scriptstyle n-l\ &\scriptstyle r-n&\scriptstyle
\nu-r\ &\scriptstyle l\ \ \ &\scriptstyle n-l\ &\scriptstyle r-n
&\scriptstyle \nu-r\ &\scriptstyle 1&\scriptstyle 1\end{array}
\end{array},$$
where $H_3$ and $H_9$ arbitrarily. Therefore, the number of $e_T$
which contained in $m$ and containing $e_L$ is $q^{(r-n+1)(n-l)}.$

(2) Similarly, by computation, we can proof that the number of $e_T$
contained in $m$ and containing $e_L,e_{R_{i}}$ has the following
the form
$$\begin {array}[t]
{cc@{\extracolsep{0.2ex}}c}e_T=\left(
\begin {array}{cccccccccccc}
I^{(l)}&0&0&0&0&0&0&0&0\ \ &0\ &0&0\\
0&I^{(n-l)}&0&0&0&0&0&0&0\ \ &0\ &0&0\\
0&0&R_3&0&I^{(l)}&0&0&0&0\ \ &0\ &R_9&0\\
0&0&H''_3&0&0&I^{(i-l-l)}&0&0&0\ \ &0\ &H''_9&0\\
0&0&H'_3&0&0&0&1&0&0\ \ &0\ &H'_9&0\\
0&0&H'''_3&0&0&0&0&I^{(n-i)}&0\ \ &0\ &H'''_9&0
\end{array} \right)\begin {array}{c}\end{array}
\\\begin {array}{cccccccccccc}\ \ \ \ \ \ \ \ \ \scriptstyle l\ \ \ \ &\scriptstyle n-l\ \ &\scriptstyle
r-n\ &\scriptstyle \nu-r&\scriptstyle l\ \ \ &\scriptstyle i-l-1\ \
\ &\scriptstyle 1\ \ &\scriptstyle n-i\ &\scriptstyle
r-n&\scriptstyle \nu-r \ &\scriptstyle 1\ \ \ &\scriptstyle
1\end{array}
\end{array},$$\\
where $R''_3, R''_9$  and $R'''_3, R'''_9$ arbitrarily. Therefore,
the number of $e_T$ contained in $m$ and containing $e_L,e_{R_{i}}$
is $q^{(n-l-1)(r-n+1)}$.

{\bf Lemma 3.6}\ Assume that $m_1$ and $m_2$ are two distinct
messages which commonly contain a transmitter's encoding rule $e_T$.
$s_1$ and $s_2$ contained in $m_1$ and $m_2$ are two source states,
respectively. Assume that $s_0=s_1\cap s_2$, dim $s_0=k$, then
$n\leq k\leq 2r-n$. For any $e_{L},e_{R_{i}}\subset m_1\cap m_2$,
the number of $e_T$ contained in $m_1\cap m_2$ and containing
$e_L,e_{R_{i}}$ is $q^{k(n-l-1)}$.

\textbf{\emph{Proof.}}\ \ Since $m_1=s_1+e_T, m_2=s_2+e_T$ and
$m_1\neq m_2$, then $s_1\neq  s_2$. For any $s \in S$, $U \in
s$,Obviously, $n\leq k\leq 2r-n$. Assume that $s'_i$ is the
complementary subspace of $s_0$ in the $s_i$, then $s_i=s_0+s'_i\,
\,\,(i=1,2)$. From $m_i=s_i+e_T=s_0+s'_i+e_T$, we have $m_1\cap
m_2=s_{0}+ e_{T}$.

From the definition of the message, we may take $m_{i},i=1,2$ as
follows
$$\begin {array}[t] {cc@{\extracolsep{0.2ex}}c}m_i=\left(
\begin {array}{cccccccccccc}
I^{(l)}&0&0\ \ &0\ &0&0&0&0\ &0&0\\
0&I^{(n-l)}&0\ \ &0\ &0&0&0&0\ &0&0\\
0&0&P_{i_3}\ \ &0\ &0&0&0&0\ &0&0\\
0&0&0\ \ &0\ &I^{(l)}&0&0&0\ \ &0&0\\
0&0&0\ \ &0\ &0&I^{(n-l)}&0&0\ &0&0\\
0&0&0\ \ &0\ &0&0&I^{(r-n)}&0\ &0&0\\
0&0&0\ \ &0\ &0&0&0&0\ &1&0
\end{array} \right)&\begin {array}{l}
\scriptstyle l\\ \scriptstyle n-l\\ \scriptstyle r-n\\ \scriptstyle
l\\ \scriptstyle n-l\\ \scriptstyle r-n\\ \scriptstyle 1
\end{array}\\&
\\\begin {array}{cccccccccc}\ \ \ \ \ \ \ \ \ \ \ \ \scriptstyle l\ \ \ \ &\scriptstyle n-l\ \ &\scriptstyle r-n\ &\scriptstyle
\nu-r\ &\scriptstyle l\ \ \ &\scriptstyle n-l\ \ \ \ &\scriptstyle
r-n &\scriptstyle \ \ \nu-r\ &\scriptstyle 1&\scriptstyle
1\end{array}
\end{array}.$$\\
Let $$\begin {array}[t] {cc@{\extracolsep{0.2ex}}c}m_1\cap
m_2=\left(
\begin {array}{cccccccccccc}
I^{(l)}&0&0\ &0&0&0&0\ \ &0\ \ &0&0\\
0&I^{(n-l)}&0\ &0&0&0&0\ \ &0\ \ &0&0\\
0&0&P_{3}\ &0&0&0&0\ \ &0\ \ &0&0\\
0&0&0\ &0&I^{(l)}&0&0\ &0\ \ &0&0\\
0&0&0\ &0&0&I^{(n-l)}&0\ \ &0\ \ &0&0\\
0&0&0\ &0&0&0&I^{(r-n)}\ \ &0\ \ &0&0\\
0&0&0\ &0&0&0&0\ \ &0\ \ &1&0
\end{array} \right)&\begin {array}{l}
\scriptstyle l\\ \scriptstyle n-l\\ \scriptstyle r-n\\ \scriptstyle
l\\ \scriptstyle n-l\\ \scriptstyle r-n\\ \scriptstyle 1
\end{array}\\&
\\\begin {array}{cccccccccc}\ \ \ \ \ \ \ \ \ \ \ \ \ \ \ \ \ \ \ \ \scriptstyle l\ \ \ &\scriptstyle n-l\ \ &\scriptstyle r-n\ &\scriptstyle\nu-r&\scriptstyle l\ \ \
&\scriptstyle n-l\ \ \ &\scriptstyle r-n \ \ \ \ &\scriptstyle
\nu-r\ &\scriptstyle 1\ &\scriptstyle 1\end{array}
\end{array}.$$\\
From above we know that $m_1\cap m_2=s_{0}+ e_{T}$, then
$dim(m_1\cap m_2)=k+2n-n=k+n$, therefore,

$$dim\left(
\begin {array}{ccccc}
P_{3}&0\ \ &0\ &0&0\\
0&0\ \ &0\ &1&0
\end{array} \right)\begin {array}{c}\end{array}
=k+n-(2n+r-n)=k-r.$$\\ For any $e_L,e_{R_i}\subset m_1\cap m_2$, we
can assume that\\
$$\begin {array}[t] {cc@{\extracolsep{0.2ex}}c}e_L=\left(
\begin {array}{cccccccccccc}
I^{(l)}&0&0&0&0&0\ \ &0\ &0\ &0\ \ &0&0&0\\
0&I^{(n-l)}&0&0&0&0\ \ &0\ &0\ &0\ \ &0&0&0\\
0&0&R_3&0&I^{(l)}&0\ \ &0\ &0\ &0\ \ &0&R_{11}&0
\end{array} \right)&\begin {array}{l}
\scriptstyle l\\ \scriptstyle n-l\\ \scriptstyle l
\end{array}\\&
\\\begin {array}{cccccccccccc}\ \ \ \ \ \ \ \ \ \scriptstyle l\ \ &\scriptstyle n-l\ \ &\scriptstyle
r-n&\scriptstyle \nu-r&\scriptstyle l&\scriptstyle
i-l-1&\scriptstyle 1&\scriptstyle n-i&\scriptstyle r-n&\scriptstyle
\nu-r &\scriptstyle 1\ &\scriptstyle 1\end{array}
\end{array},$$\\
$$\begin {array}[t] {cc@{\extracolsep{0.2ex}}c}e_{R_{i}}=\left(
\begin {array}{cccccccccccc}
I^{(l)}&0&0&0\ &0\ &0\ \ &0\ &0\ &0\ \ &0\ &0&0\\
0&I^{(n-l)}&0&0\ &0\ &0\ \ &0\ &0\ &0\ \ &0\ &0&0\\
0&0&H'_3&0\ &0\ &0\ \ &1\ &0\ &0\ \ &0\ &H'_{11}&0
\end{array} \right)&\begin {array}{l}
\scriptstyle l\\ \scriptstyle n-l\\ \scriptstyle 1
\end{array}\\&
\\\begin {array}{cccccccccccc}\ \ \ \ \ \ \ \ \ \ \scriptstyle l\ \ &\scriptstyle n-l\ \ &\scriptstyle
r-n&\scriptstyle \nu-r&\scriptstyle l&\scriptstyle
i-l-1&\scriptstyle 1&\scriptstyle n-i&\scriptstyle r-n&\scriptstyle
\nu-r \ &\scriptstyle 1\ \ &\scriptstyle 1\end{array}
\end{array},$$\\
If $e_T \subset m_1\cap m_2$ and containing $e_L, e_{R_{i}}$, so
$e_T$ has the form as follows\\
$$\begin {array}[t] {cc@{\extracolsep{0.2ex}}c}e_T=\left(
\begin {array}{cccccccccccc}
I^{(l)}&0&0&0&0&0&0&0&0\ \ &0\ &0&0\\
0&I^{(n-l)}&0&0&0&0&0&0&0\ \ &0\ &0&0\\
0&0&R_3&0&I^{(l)}&0&0&0&0\ \ &0\ &R_{11}&0\\
0&0&H''_3&0&0&I^{(i-l-l)}&0&0&0\ \ &0\ &H''_{11}&0\\
0&0&H'_3&0&0&0&1&0&0\ \ &0\ &H'_{11}&0\\
0&0&H'''_3&0&0&0&0&I^{(n-i)}&0\ \ &0\ &H'''_{11}&0
\end{array} \right)&\begin {array}{l}
\scriptstyle l\\ \scriptstyle n-l\\ \scriptstyle l\\ \scriptstyle
i-l-l\\ \scriptstyle 1\\ \scriptstyle n-i
\end{array}\\&
\\\begin {array}{cccccccccccc}\ \ \ \ \ \ \ \ \ \scriptstyle l\ \ \ \ &\scriptstyle n-l\ \ &\scriptstyle
r-n\ &\scriptstyle \nu-r&\scriptstyle l\ \ \ &\scriptstyle i-l-1\ \
\ &\scriptstyle 1\ \ &\scriptstyle n-i\ &\scriptstyle
r-n&\scriptstyle \nu-r \ &\scriptstyle 1\ \ \ &\scriptstyle
1\end{array}
\end{array}.$$\\
where every row of
$$\begin {array}[t] {cc@{\extracolsep{0.2ex}}c}\left(
  \begin {array}{ccccc}
R''_3&0&0&R''_{11}&0\\
R'''_3&0&0&R'''_{11}&0
\end{array} \right)\begin {array}{c}\end{array}
   \\\begin {array}{cccccccccc}\ \ \ \ \ \ \ \ \ \ \ \ \ \ \ \scriptstyle \ \ &\scriptstyle &\scriptstyle
  \ &\scriptstyle \ \ \ \ &\scriptstyle \ \ &\scriptstyle \ \ &\scriptstyle \ \ &\scriptstyle &\scriptstyle
  \ &\scriptstyle \end{array}
  \end{array}$$\\
is the linear combination of the base of\\ $$\begin {array}[t]
{cc@{\extracolsep{0.2ex}}c}\left(
\begin {array}{ccccc}
P_{3}&0\ \ &0\ &0&0\\
0&0\ \ &0\ &1&0
\end{array} \right).\begin {array}{c}\end{array}
   \\\begin {array}{cccccc}\ \ \ \ \scriptstyle \ \ \ \ \ \ \ &\scriptstyle  \ \ \ \ \ \ &\scriptstyle
  \ \ \ \ \ \ \ &\scriptstyle\ \ \ \ &\scriptstyle \ \ \ &\scriptstyle \end{array}
  \end{array}$$
So it is easy to know that the number of $e_T \subset m_1\cap m_2$
and containing $e_L, e_{R_{i}}$ is $q^{(k-r)(n-l-1)}$.\\

{\bf Theorem  3.7} \ In the constructed multi-receiver
authentication codes, the largest probabilities of success for
$impersonation\ attack$ and $substitution\ attack$ from $R_{L}$ on a
receiver $R_{i}$ are
$$P_{I}[i,
L]=\frac{1}{q^{(n-l)(\nu-r)+(r-n+1)}}, \ \ \ \ \ \ \ \ \\
P_{S}[i, L]=\frac{1}{q^{r-l}}$$ respectively, where $i\notin L$.

{\bf \emph{Proof.}}\ \ $Impersonation\ attack$: $R_L$, after receiving
their secret keys, send a message $m$ to $R_i$. $R_L$ is successful
if $m$ is accepted by $R_i$ as authentic. Therefore
\begin{eqnarray}P_{I}[i, L]
&=&\max\limits_{e_{L}\in E_{L}}\left\{\frac{\max\limits_{m\in
M}\mid\{e_{T}\in E_{T}|e_{T}\subset m {\rm \ and} \ e_{T}\supset
e_{L},e_{R_i}\}\mid}{\mid\{e_{T} \in E_{T}|e_{T}\supset
e_{L}\}\mid}\right\}\nonumber\\
&=&\frac{q^{(n-l-1)(r-n+1)}}{q^{(\nu-n+1)(n-l)}}\nonumber\\
&=&\frac{1}{q^{(n-l)(\nu-r)+(r-n+1)}}.\nonumber\end{eqnarray}

$Substitution\ attack$: $R_L$, after observing a message $m$ that is
transmitted by the sender, replace $m$ with another message $m'$.
$R_L$ is successful if $m'$ is accepted by $R_i$ as authentic.
Therefore
\begin{eqnarray}P_{S}[i, L]&=&\max\limits_{e_{L}\in E_{L}}\max\limits_{m\in
M}\left\{\frac{\max\limits_{m'\in M}\mid\{e_{T}\in
E_{T}|e_{T}\subset m,m' {\rm \ and} \ e_{T}\supset
e_{L},e_{R_i}\}\mid}{\mid\{e_{T}\in E_{T}|e_{T}\subset m{\rm \ and}
\ e_{T}\supset e_{L}\}\mid}\right\}\nonumber\\
&=&\max\limits_{n\leq k\leq 2r-n}\frac{q^{(k-r)(n-l-1)}}{q^{(n-l)(r-n+1)}}\nonumber\\
&=&\frac{1}{q^{r-l}}. \nonumber\end{eqnarray}

From above we see, $substitution\ attack$ from $R_{L}$ on a receiver
gets to the maximum when $l=r-1$.

\end{document}